\documentclass[
superscriptaddress,
 amsmath,amssymb,
 prl,
 aps,
 twocolumn
]{revtex4-1}
\usepackage{graphicx}
\usepackage{dcolumn}
\usepackage{bm}
\usepackage{natbib}
\usepackage{pgfplots}
\usepackage[utf8]{inputenc}  
\usepackage[T1]{fontenc}     
\usepackage{hyperref}
\usepackage{tabularx}
\usepackage{wrapfig}
\usepackage{nameref}

\definecolor{royalred}{rgb}{0,0,0}
\definecolor{royalred}{rgb}{0.6078,0.1098,0.1922}
\definecolor{darkblue}{rgb}{0.3059 , 0.4471, 1}
\definecolor{darkred}{rgb}{0.9098 , 0.2824, 0.3333}
\definecolor{darkgreen}{rgb}{0.3176 , 0.5961, 0.4471}
\definecolor{bleu}{rgb}{0.1,0.2,0.6}

\begin{document}

\title{Multiscale interfacial mechanics of soft solids}

\author{Nicolas Bain}
 \email{nicolas.bain@cnrs.fr}
 \affiliation{Universite Lyon 1, CNRS, Institut Lumière Matière, UMR5306, F-69100, Villeurbanne, France}
 \affiliation{Department of Materials, ETH Z\"{u}rich, 8093 Z\"{u}rich, Switzerland.}

  \author{Lawrence A. Wilen}
 \affiliation{Mechanical Engineering and Materials Science, Yale University, New Haven, CT, USA.}

 \author{Dominic Gerber}
 \affiliation{Department of Materials, ETH Z\"{u}rich, 8093 Z\"{u}rich, Switzerland.}

 \author{Mengjie Zu}
 \affiliation{Institute of Science and Technology Austria (ISTA), Am Campus 1, 3400 Klosterneuburg, Austria.}

 \author{Carl P. Goodrich}
 \affiliation{Institute of Science and Technology Austria (ISTA), Am Campus 1, 3400 Klosterneuburg, Austria.}

 \author{Senthilkumar Duraivel}
 \affiliation{Department of Materials Science and Engineering, Cornell University, Ithaca, NY, USA.}

 \author{Kaarthik Varma}
 \affiliation{Department of Physics, Cornell University, Ithaca, NY, USA.}

 \author{Harsha Koganti}
 \affiliation{Department of Materials Science and Engineering, Cornell University, Ithaca, NY, USA.}

 \author{Robert W. Style}
 \affiliation{Department of Materials, ETH Z\"{u}rich, 8093 Z\"{u}rich, Switzerland.}

 \author{Eric R. Dufresne}
 \affiliation{Department of Materials, ETH Z\"{u}rich, 8093 Z\"{u}rich, Switzerland.}
 \affiliation{Department of Materials Science and Engineering, Cornell University, Ithaca, NY, USA.}
 \affiliation{Department of Physics, Cornell University, Ithaca, NY, USA.}

\date{\today}

\begin{abstract}
\textbf{
\noindent
Soft solids and their surface deformations control the response of many natural and artificial systems.
Yet, their underlying properties are vigorously debated, particularly for polymer networks.
While molecular-scale theories predict no interfacial changes with macroscopic deformation, multiple experiments suggest otherwise.
To settle this issue, we measure displacement fields near the interface of a silicone gel, in the limit of small deformations.
We discover an unexpected multiscale response.
The shear modulus decreases smoothly by half with 20 $\mu$m of the interface.
At the same time we observe a surface excess elasticity, that depends on history and outer medium composition.
These results reveal the fundamentally multiscale nature of polymeric surfaces, and call for further experimental and theoretical investigations into the basic understanding of soft solid interfaces.
}
\end{abstract}

\pacs{Valid PACS appear here}
\maketitle

\section{Introduction}
\label{sec:intro}

\noindent
Interfaces play a crucial role in mechanics and materials.
They are central to composites and robotics \cite{baughman2002carbon,style2015stiffening, ortega2025ultrafast,koh2015jumping}, adhesion and friction \cite{karpitschka2016surface,hui2020surface,saintyves2016self, lee2008sweet,creton2003pressure}, fracture \cite{bostwick2013capillary,liu2019elastocapillary,creton2016fracture}, wetting \cite{de1985wetting, duprat2012wetting, bardall2018deformation,jerison2011deformation,cai2021fluid}, shape morphing \cite{bico2018elastocapillarity, paretkar2014flattening, mora2013solid}, and mechanobiological processes in cells and tissues \cite{ehrig2019surface, shi2022elastocapillarity, yousafzai2022cell}.
For solids, interfacial tension can play a dominant role at length scales smaller than the \emph{elastocapillary length}, equal to the ratio of interfacial tension, $\Upsilon$ and shear modulus, $\mu$ \cite{style2017elastocapillarity,andreotti2020statics, chen2018static}.
This cross-over is widely accepted, but the fundamental nature of interfacial tension for soft polymeric solids is still controversial.
In theory, ideal polymer chains should seamlessly rearrange at a deformed solid interface, leading to a strain-independent interfacial tension \cite{liang2018surface, liang2018surface}.
In practice, observations of microscopic deformations in silicone gels suggest a strain-dependent interfacial tension \cite{xu2017direct,jensen2017strain,xu2018surface,bain2021surface, bardall2018deformation}.
The interpretation of these previous experiments, however, is complicated by material and geometric nonlinearities \cite{masurel2019elastocapillary,pandey2020singular,heyden2022robust,heyden2024distance}.

In this paper, we quantify mechanical properties near the interface of a silicone gel with high precision 3D location and tracking (\hyperref[sec:qd_appendix]{Appendix A}).  We apply homogeneous or localized stresses, and analyze deformation fields in the linear response regime.
First, we find a two-fold drop of the shear modulus, within 20 $\mu$m of the surface.
Second, we find a surface excess elastic response, at unresolved scales, that depends on the composition of the surrounding medium and on deformation history.
These results demonstrate the multiscale nature of soft polymeric solid interfaces, in contrast with the scale-free elastic models that are used widely to infer traction stresses or bulk elastic moduli \cite{radmacher1992molecules,style2014traction,cheung20243d,norman2021measuring,rieu2005direct,delanoe20104d}.
We thus need new approaches to understand, model, and design interfaces that account for their multiple intrinsic length scales.

\section{Experimental Design}

\begin{figure}[b!]
\includegraphics[scale = 1]{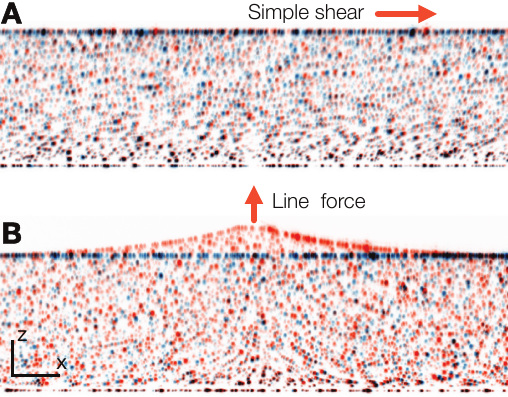}
\vspace{-10pt}
\caption{
Superimposed max-intensity projections of confocal stacks, before deformation (blue) and after deformation (orange), in two deformation modes (fraction of data - 10\%).
(A) Simple shear.
(B) Line force.
The scale bars defining the axis each equate 10\,$\mu$m. 
The black dots result from the superimposition of blue and orange at the same location.
}
\vspace{-15pt}
\label{fig_sketch}
\end{figure}

\begin{figure*}[t]
\includegraphics[scale = 1]{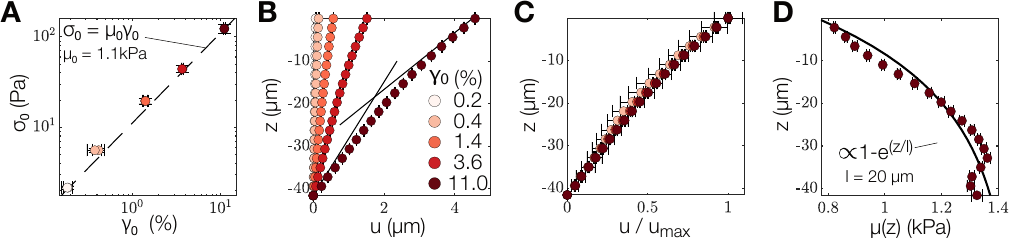}
\vspace{-10pt}
\caption{
(A) Bulk strain-stress relationship, together with the line $\sigma_0 = \mu_0\gamma_0$ with $\mu_0 = 1.1$kPa.
Errors bars correspond to twice the standard deviation.
(B) Horizontal displacements $u$ against the vertical position $z$.
The black lines correspond to a linear fit of the first two and the last two dark red data points.
(C) Horizontal displacements normalized by the maximal horizontal displacement $u_{\rm max}$. 
(D) Local shear modulus $\mu(z)$ against the vertical position $z$.
Each color codes for an imposed macroscopic stress.
Plots (B)-(D) correspond to values averaged over 2.2 $\mu$m thick horizontal bins. The size of the errorbars equates twice the standard error of the mean (\hyperref[sec:err]{Appendix D}). 
}
\vspace{-15pt}
\label{fig_bulk}
\end{figure*}

We deform films of soft silicone gels in two geometries.
First, we apply a simple shear stress (Fig.~\ref{fig_sketch}A) with a magnet adhered to the top of the film (\hyperref[sec:shear]{Appendix B}).
This creates a homogeneous stress throughout the material.
In this way, nonlinearities in strain profiles reveal material heterogeneities.
Second, we apply a localized force per unit length, using the three-phase contact line of a liquid droplet (Fig.~\ref{fig_sketch}B) or from the edge an adhered coverslip (\hyperref[sec:coverslip]{Appendix C}). 
In the latter experiments, observation of shear stresses at the interface reveals strain-dependent interfacial tension.
In each case, we precisely measure the deformation fields with confocal microscopy of embedded quantum dots (\hyperref[sec:supp_videos]{Supplementary Video 1}). 
The small size of these fluorescent tracers allows us to densely sample the deformation field, with an average spacing of $3~\mu\mathrm{m}$, at vanishingly small volume fractions ($10^{-5}$).
This allows us to quantify strains as small as $0.2\%$ in $x$,$y$ and $0.4\%$ in $z$ (\hyperref[sub:polyharmonic]{Appendix A.4}).

\section{Simple Shear reveals near-surface gradients in bulk elasticity}

We measure three-dimensional deformation fields across the thin silicone film, as we increase the applied shear stress with a calibrated magnet-based setup (Fig.~\ref{fig_sketch}A, \hyperref[sec:shear]{Appendix B}).
The average shear strain across the film is $\gamma_{0} = [u(h)-u(0)]/h$, where $u$ is the horizontal displacement and $h$ the thickness of the sample.
It increases linearly with the applied stress $\sigma_0 = \mu_0 \gamma_0$ (Fig.~\ref{fig_bulk}A), with a shear modulus of $\mu_0 = 1.1$ kPa consistent with previous experiments on this material \cite{bain2021surface}.

We now investigate the local mechanical response through analysis of  the full three-dimensional displacement field.
As expected, we find that displacements only depend on $z$, and are independent of $x$ and $y$ (\hyperref[sec:z_dep_disp]{Appendix B.3}).
The $z$-dependence of the displacements are shown in Fig.~\ref{fig_bulk}B.
Instead of the linear profile expected for a homogeneous material, we observe that the horizontal motion of the tracers decays non-linearly, faster close to the top surface (Fig.~\ref{fig_bulk}B, \hyperref[sec:shear_non_lin]{Appendix B.4}).
This behavior is independent of strain, over the range of measured values from $0.4\%$ to $11.0\%$, as shown by the collapse of  displacement profiles on a single curve when normalized by the maximal displacement (Fig.~\ref{fig_bulk}C).

Since force balance requires the stress to be homogeneous, the non-linear displacement profile implies a heterogeneous shear modulus, $\mu(z) = \sigma_0 / \gamma(z)$, with $\gamma(z) = \partial u / \partial z$.
The depth dependence of the shear modulus is shown in Fig. \ref{fig_bulk}D.
The shear modulus nearly doubles, from $765 \pm 3 \mbox{ Pa}$ at the top surface, to $1320 \pm 20 \mbox{ Pa}$ at the bottom surface.
Its  profile is in reasonable agreement with an exponential decay, with decay length $l = 20\,\mu\mbox{m}$ (Fig.~\ref{fig_bulk}D).
This gradient of shear modulus suggests a gradient in the network crosslinking density.
Its profile is consistent with a reaction-diffusion process \cite{turing1990chemical,kondo2010reaction}, 
 such as oxygen-inhibition of cross-linking 
\cite{mandal2021oxygen}.
The precise origin of this heterogeneity is however unknown and left for future investigation (\hyperref[sec:mech_hetero]{Appendix B.5}).

\begin{figure*}[t]
\includegraphics[scale = 1]{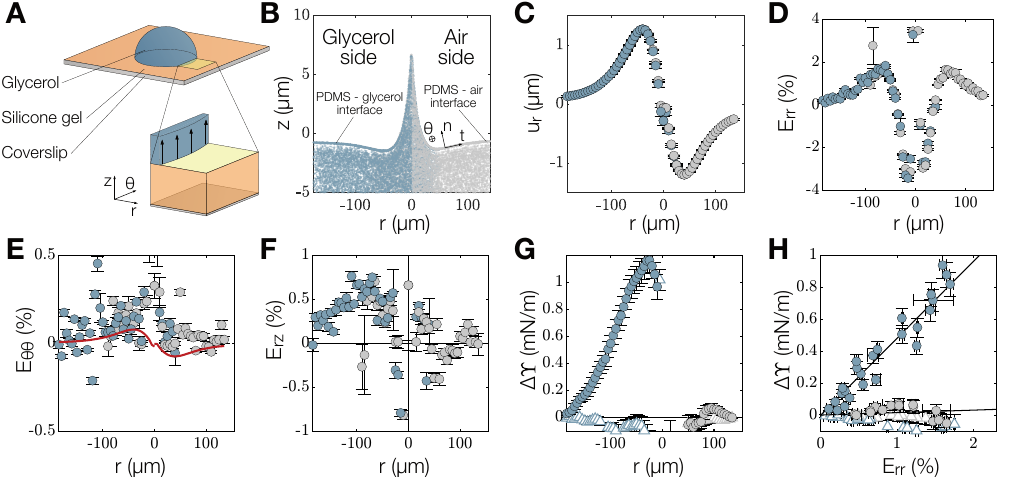}
\vspace{-5pt}
\caption{
(A) Sketch of the three-phase contact line confocal imaging. We acquire confocal stacks of both the inner and the outer side of the contact line.
Sketches not to scale.
(B) Azimuthally-collapsed tracer positions on both sides of the wetting ridge.
Surface-deposited tracers are distinguishable from the bulk-included quantum dots.
$\mathbf t$, $\mathbf n$, and $\bm \theta$ represent the unit vectors respectively tangent, normal to the surface, and in the orthoradial direction.
(C) Surface radial displacement $u_{r}$.
(D), (E), and (F) 
Surface radial strain $E_{\rm rr}$, orthoradial strain $E_{\theta\theta}$, and shear strain $E_{\rm rz}$.
(E), the red curve corresponds to the expected orthoradial component $u_r / R$.
(G) and (H) Excess interfacial tension $\Delta \Upsilon$ against the radial position $r$, and against the surface strain $E_{\rm rr}$.
In (H), the solid lines correspond to $\Delta\Upsilon = \Lambda E_{\rm rr}$, with $\Lambda = 1.6$~mN/m, and $\Lambda = 48.1$~mN/m.
In (C) to (H), the symbols correspond to values averaged over 5 $\mu$m wide radial bins, and the size of the errorbars equate twice the standard error of the mean (\hyperref[sec:err]{Appendix D}).
The filled circles correspond to measured values, and the non-filled triangles in (G) and (H) to bulk-extrapolated values  . 
In (B) to (H), the blue and gray colors correspond to confocal datasets with most  of the field of view on the glycerol and air sides, respectively.
}
\vspace{-10pt}
\label{fig_droplet}
\end{figure*}

\section{Force Balance at Interfaces}

The net force on any part of a material at rest must equal zero.
For volume elements spanning the interface, this force balance includes the bulk elastic stress $\bm \sigma$ and the interfacial excess stress $\Upsilon$.
When applied to the normal forces, we arrive at the familiar Laplace equation: $\mathbf n \cdot \bm \sigma \cdot \mathbf n = \sigma_{\rm nn}= \kappa\Upsilon$, where $\kappa$ is the local curvature, and $\mathbf n$ the unit vector normal to the surface.
When applied in-plane, the shear stress is balanced by the gradient of interfacial stress: $\mathbf t \cdot \bm \sigma \cdot \mathbf n = \sigma_{\rm tn} = \partial \Upsilon / \partial s$, where $\mathbf t$ is the unit vector tangent to the surface, and $s$ is the arc-length material coordinate \cite{style2017elastocapillarity,pandey2020singular}.
In liquids, this is called the Marangoni stress \cite{scriven1960marangoni}.
Therefore, the shear stress must be non-zero when interfacial tension is  heterogeneous.
In the following, we extrapolate bulk stresses to the interface, and identify the interfacial tension gradient $\partial \Upsilon / \partial s$ as the tangential component of the unbalanced elastic stress.

\section{Stress gradients unveil interface elasticity}

We measure  three-dimensional deformation fields near the three phase contact line of a glycerol droplet resting on a thin silicone film (Figs.~\ref{fig_sketch}B, ~\ref{fig_droplet}A, \hyperref[sec:ridge_ref_state]{Appendix E.1}).  
The interfacial tension of the liquid-air interface induces large deformations near the contact line \cite{style2013universal}.
Azimuthally-collapsed tracer positions near the contact line in are shown in Fig.~\ref{fig_droplet}B, and the radial displacement $u_{\rm r}$ of the surface in Fig.~\ref{fig_droplet}C (see \hyperref[sec:ridge_profile]{Appendix E.2} for a 3D profile and a closer look at the ridge shape). 
We made sure the contact line is fully equilibrated when we conduct the imaging (\hyperref[sec:ridge_profile]{Appendix E.2}).
Previous analyses of these profiles, which focused on the structure of the tip of the wetting ridge \cite{style2013universal}, were vulnerable to elastic singularities and solvent extraction \cite{flapper2023reversal,pandey2020singular,cai2021fluid,masurel2019elastocapillary}.
Here, we instead focus our analysis \emph{far from the contact line}, where the deformations remain small. 

To determine the bulk elastic stresses at the interface, we calculate the deformation gradient tensor $\mathbf F = \mathbf I + \bm\nabla\mathbf u$, and assume compressible Neo-Hookean behavior $\bm\sigma = \mu(z)J^{-1}\mathbf F\mathbf F^{\text{\tiny T}} - p\mathbf I$, where $p$ is a pressure term and $J = \det\mathbf F$ accounts for local volume changes \cite{anand2023introduction}.
To estimate the strains at the surface, we calculate the Green strain $\mathbf E = (\mathbf F^{\text{\tiny T}}\mathbf F - \mathbf I) /2$.
The surface radial strain, $E_{\rm rr} = \hat{\mathbf r}\cdot\mathbf E\cdot \hat{\mathbf r}$ with $\hat{\mathbf r}$ the unit vector in the radial direction, is shown in Fig.~\ref{fig_droplet}D.
The profile is roughly symmetric about the contact line. 
Within $100\,\mu\mbox{m}$ of the contact line,  the strain is  relatively large and quickly varying.
Further away, it decays smoothly to zero.
The strain in the orthoradial direction, $E_{\theta\theta} = \hat{\bm \theta}\cdot\mathbf E\cdot\hat{\bm \theta}$ with $\hat{\bm \theta}$ the unit orthoradial vector, is shown in Fig.~\ref{fig_droplet}E.
It is much smaller than the strain in the radial direction.
As all displacements only depend on $r$ and $z$, the magnitude of $E_{\theta\theta}$ is comparable to $u_{\rm r} / R$, shown in red, where $u_{\rm r}$ is the radial displacement and $R$ the droplet radius \cite{landau1960mechanics}. 
Last, the shear strain $E_{\rm rz} = \hat{\mathbf r}\cdot\mathbf E\cdot\hat{\mathbf z}$ with $\hat{\mathbf z}$ the unit vertical vector, shown in Fig.~\ref{fig_droplet}F, is asymetric about the contact line.
For sake of completeness, we report the surface normal and volumetric strains in \hyperref[sec:surf_nom_vol]{Appendix E.4}.
We note that the bulk volumetric strains, reported in \hyperref[sec:vol_bulk]{Appendix E.5}, are indicative of a net poroelastic response with a surprisingly rich spatial structure.

Applying the Marangoni boundary condition, we calculate the incremental interfacial tension $\Delta \Upsilon = \int \sigma_{\rm tn} \mbox{d}s$ on either side of the contact line by integrating the shear strain from the far field along the arc-length $s$ (Fig.~\ref{fig_droplet}G) \cite{pandey2020singular}.
There, $\sigma_{\rm tn} = \mathbf t\cdot\bm\sigma\cdot\mathbf n$ is the projection of the surface stress onto $\mathbf t$, the unit vector tangent to the surface pointing in the radial direction, and $\mathbf n$ the unit vector normal to the surface (Fig.~\ref{fig_droplet}B, \hyperref[sec:tang_norm]{Appendix E.3}), using the shear modulus evaluated at the surface $\mu(h).$

At the PDMS-glycerol interface, the shear strain $E_{\rm rz}$ is positive (Fig.~\ref{fig_droplet}F).
As the shear stress $\sigma_{\rm tn}$ is at first order proportional to the shear strain $E_{\rm rz}$, the incremental tension increases as the contact line is approached (Fig.~\ref{fig_droplet}G).
When compared to the tangential strain $E_{\rm rr}$, we find a linear increase in incremental tension (Fig.~\ref{fig_droplet}H).
This suggests a linear elastic response of the interface: $\Delta\Upsilon = \Lambda E_{\rm rr}$ with a surface modulus $\Lambda=48.1\pm 4.4~\mathrm{mN/m}$.
In contrast, the shear strain on the PDMS-air side fluctuates about zero (Fig.~\ref{fig_droplet}F).
The incremental tension therefore does not significantly deviate from zero (Fig.~\ref{fig_droplet}G), and when compared to the surface strain, it gives $\Lambda=1.6 \pm 2.2~\mathrm{mN/m}$.
Thus, the interfacial tension of the PDMS-air interface is effectively  independent of surface strain.
The contrast in interface elasticity across the contact line was observed consistently across five different instances (\hyperref[sec:surf_el_all_samples]{Appendix E.6}), and complemented with additional measurements on the deformed PDMS-air interface at the edge of the coverslip (\hyperref[sec:edge]{Appendix F}).

\section{Displacement jump reveals surface alteration}

\begin{figure}[t]
\includegraphics[scale = 1]{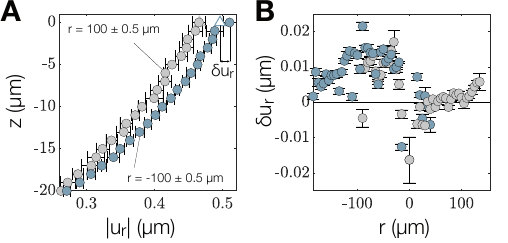}
\vspace{-10pt}
\caption{
Radial displacement surface jump.
(A) Radial absolute displacement profile, at $-100\,\mu\mbox{m} \pm 0.5\,\mu\mbox{m}$ (blue) and $100\,\mu\mbox{m} \pm 0.5\,\mu\mbox{m}$ (gray) from the position of the wetting ridge, against vertical position $z$.
Each symbol corresponds to values averaged over $1 \mu\mbox{m} \times 1 \mu\mbox{m}$ r-z bins.
The circles correspond to measured displacements, and the triangle to bulk-extrapolated displacements.
(B) Radial displacement jump against radial position $r$, same color-code as in Fig.~\ref{fig_droplet}.
Each circle corresponds to surface values averaged over 5 $\mu$m wide radial bins.
In (A) and (B), the size of the errorbars equate twice the standard error of the mean (\hyperref[sec:err]{Appendix D}). 
}
\vspace{-15pt}
\label{fig_jump}
\end{figure}
To isolate the features of the data that encode surface elasticity, we examine the radial displacement vertical profile $u_{\rm r}(z)$ (Fig.~\ref{fig_jump}A).
On the air side, where no surface elasticity is observed, $u_{\rm r}(z)$ increases smoothly from bottom to top.
On the glycerol side, where we observed significant surface elasticity, $u_{\rm r}(z)$ jumps when reaching the top surface.
This discontinuity occurs over a sub-micron length scale, much smaller than the elasticity gradient described above, and differs from previously reported interface slippage in which bulk and surface move as one \cite{newby1995macroscopic,zhang1997effect}.
We quantify the jump amplitude, $\delta u_{\rm r}$, as the difference between measured and bulk-extrapolated surface displacements (Fig.~\ref{fig_jump}B).
We find it to be of the order of $10\,\mbox{nm}$ along the PDMS-glycerol interface, and close to $0\,\mbox{nm}$ along the PDMS-air interface.

We test the significance of the displacement jump by reanalyzing the data excluding tracers attached to the surface interfacial tension (\hyperref[sec:bulk_extr]{Appendix G}).
In this case, interfacial tension appears to be independent of surface strain on both sides of the interface (Fig.~\ref{fig_droplet}G and H).
In other words, the observed displacement jump at the top surface is responsible for the measured interfacial elasticity.
In turn, the precision of the measured interfacial elasticity is limited by the spatial resolution of the surface discontinuity.
We report this systematic behavior in the other samples we tested in \hyperref[sec:disp_jump_app]{Appendix H}.

Among the phenomena that might lead to observed differences in the surface response on the two sides of the interface, we verified that the shear stress on the glycerol side could neither be explained by an optical artifact (\hyperref[sec:dis_jump_not_artifact]{Appendix I}), nor by hydrodynamic flows (\hyperref[sec:supp_videos]{Supplementary Video 2}).
We are thus left with two hypotheses.
First, molecular interactions between glycerol and PDMS.
On the one hand, we estimated that glycerol swells this silicone gel by less than 3ppm (\hyperref[sec:glycerol]{Appendix J}), making localized changes to material properties due to the dissolution of glycerol into polymer network unlikely \cite{lee2003solvent, chen2019distinguishing}.
On the other hand, recent experiments have shown that surfacing PDMS chains could restructure in contact to inert solvents such as water or glycerol \cite{chen2004surface,park2012surface,nannette2024thin}.
Second, mechanical damage.
As the glycerol droplet is grown to its final shape, the PDMS undergoes large deformation at the wetting ridge and permanent damage could alter the mechanical behavior over a thin layer \cite{thanawala2000surface}.
This hypothesis would be consistent with recent surface elasticity measurements on a soft PDMS after it was peeled off from a patterned mold \cite{bain2021surface}.

Whether the surface apparent surface elasticity has a chemical or a mechanical origin calls for further investigations.
These results however demonstrate that surface elasticity can be tuned independently of bulk properties by the mere passage of a liquid drop,  highlighting the fundamental nature of interface elasticity as an interfacial property and opening the way for its direct engineering.

\section{Discussion and Conclusion}

We measured  bulk and interface properties of a polymer gel through high-precision tracking of nanotracers under controlled deformations. 
We resolved a gradient of the shear modulus within 20 $\mu$m of the interface, and quantified an environment- and history- dependent interface excess elasticity that arises from a mechanical discontinuity over an unresolved length scale.

These results suggest a hierarchy of length scales associated with the interface, and invite us to think carefully about the definition of interfacial properties.
Interfacial properties, as defined by Gibbs, emerge from the variation of some bulk property over an unresolved length scale near an interface \cite{gibbs1878equilibrium}. 
This definition does \emph{not} assume that interfacial excess properties emerge at the smallest structural scale,  just that they are unresolved. 
Interfacial properties can therefore arise from excess quantities spanning over molecular or supramolecular scales. 
In a polymer network, interfacial properties could be defined at the monomer scale, the mesh scale, or over larger length scales defined by network heterogeneities.

In our experiments,  depth-dependent moduli, varying over tens of microns, suggest  network heterogeneities. 
By contrast, interface elasticity emerges at an unresolved (\emph{i.e.} sub-micron) scale.
It is incompatible with the molecular view of ideal polymer networks, whose flexible chains make interfacial properties independent of deformation \cite{liang2018surface}.
Since interface elasticity is observed against glycerol, but not against air, we suspect either that mesh-bound moities are anchored to the interface in glycerol but not in air, or that large deformation at the wetting ridge resulted in surface-localized damage.
Indeed, recent experiments \cite{nannette2024thin} have shown that oxygen along the backbone of PDMS is anchored to glycerol and can prevent fusion of glycerol droplets in silicone oil, and others have shown that a PDMS surface which was peeled off a mold exhibits a net interfacial elasticity \cite{bain2021surface}.

These observations demand a fresh perspective on the interfacial properties and mechanical response of soft polymer solids.
From a polymer science perspective, we need a clear understanding of the molecular and mesh-scale features that can govern interface elasticity and better control of crosslinking to avoid or enhance larger scale network heterogeneities. From a mechanical perspective, we need clear framework for modeling materials with a multi-scale interfacial response.  
While most experiments characterizing polymer networks at the micron-scale assume a scale-free response (\emph{e.g.} Hertzian contact mechanics \cite{johnson1987contact} or traction force microscopy \cite{style2014traction,harris1980silicone}), our results show that there are at least four length scales in the problem: the elastocapillary length ($\Upsilon/\mu$), a length scale associated with interface elasticity ($\Lambda/\mu$), the length scale of network heterogeneity ($l$), and the length scale of displacement discontinuity.
In the current experiments, the three former length scales are roughly $20~\mu\mathrm{m}$, and the latter is unresolved (\emph{i.e.} sub-micron).
We expect these values to differ in other systems.
The resulting mechanical response would then qualitatively differ, depending on the relative values of all length scales at play.

\section{Author contributions:} 
N.B. and E.R.D. conceived the project and designed the experiments. 
L.A.W. designed and made the magnetic holder, and performed the magnet calibration curves.
S.D., K.V., and H.K performed the swelling and the NMR measurements.
D.G. performed the glycerol flooding measurement.
N.B. performed all the other experiments.
N.B. analyzed all the experiments.
N.B. and E.R.D. interpreted the experiments, with input from M.Z, C.P.G, and R.W.S.
N.B. and E.R.D. wrote the paper with inputs from all authors.
E.R.D. and N.B. supervised the project.
\section{Acknowledgements:} 
The authors thank 
Katharine Jensen,
Stefanie Heyden,
Thomas Salez,
Francesco Stellacci, 
Denis Bartolo,
Francesco Picella,
Hélène Delanoë-Ayari,
Mathieu Leocmach,
Antoine Bérut,
Cécile Cottin-Bizonne,
Anne-Laure Biance,
and Oriane Talabart for useful discussions.
We also thank the reviewers for excellent suggestions that substantively improved the manuscript.

\section{Appendix A: Quantum dots high precision 3D location and tracking}
\label{sec:qd_appendix}

\begin{figure*}
\includegraphics[width=\textwidth]{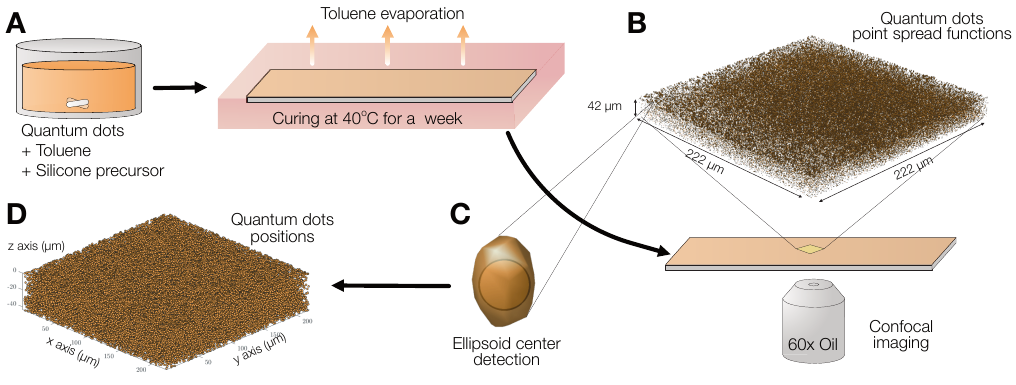}
\vspace{-10pt}
\caption{
(A) Schematic of the quantum dots incorporation process.
(B) Three-dimensional confocal reconstruction of the quantum dots point spread functions.
(C) We determine the tridimensional position of each quantum dot by fitting the associated point spread function with an ellipsoid.
(D) Three-dimensional measured positions of all quantum dots in a confocal stack.
Sketches not to scale.
}
\vspace{-15pt}
\label{fig_qd}
\end{figure*}

In this section, we detail how we incorporate quantum dots into soft polydimethylsiloxane (PDMS), detect their positions, track their displacements, estimate displacement gradients, and estimate measurement precision.

\subsection{A.1 Soft PDMS seeded with quantum dots}
\label{subs:density}

We summarize the process to seed a soft PDMS with quantum dots in Fig.~\ref{fig_qd}A.
First, we sonicate the solution of quantum dots solution (CdSeS/Zns alloyed quantum dots, Sigma Aldrich, \#753785, 1~mg per mL) for 5 minutes, and dilute 0.2 mL of this solution in 1.2 mL of toluene in a glass vial with a magnetic stirrer at 350 rpm.
Then, we mix 1.25 g of part A and 1.25 g of part B of PDMS precursor (Dowsil\texttrademark CY52-276 A\&B), and very slowly pour the mix into the glass vial.
As soon as the mixture looks homogeneous, we put a few drops of it onto a 24 mm $\times$ 50 mm  coverslip, place it in a spincoater, and spin it at 250 rpm for 60 seconds.
We then place the coverslip in a vacuum chamber and apply vacuum for 30 minutes to rapidly evaporate the excess toluene.
At last, we put the coverslip in an oven at 40$^\circ$C for a week to ensure full crosslinking of the PDMS gel.
Before imaging, we deposit a layer of 100 nm fluorescent beads following usual protocols \cite{jerison2011deformation,style2013universal,xu2017direct,xu2018surface}.

\subsection{A.2 Confocal imaging}
\label{subs:confocal}

Under confocal microscopy, the fluorescence of the quantum dots reveals their positions, which we capture in 3D image stacks (Fig.~\ref{fig_qd}B, see \hyperref[sec:supp_videos]{Supplementary Video 1} for a typical stack). 
We used a spinning disk confocal system (Yogawa CSU-X1) mounted on an inverted microscope (Nikon, Eclipse Ti2), with a 60x immersion oil objective (MRH02601), with the following settings: 488q laser power 0 and exposure time 1s (to let the immersion oil relax),
560q laser power 100 and exposure time 100ms, and z step size 0.27 $\mu$m.
To correct for the refractive index difference between the immersion oil $n_{\mbox{oil}} = 1.518$ and the silicone gel $n_{\mbox{sil}} = 1.4$ \cite{deguchi2015viscoelastic}, we rescale all z-measurements with the ratio $n_{\mbox{sil}} / n_{\mbox{oil}}$.

\subsection{A.3 3D detection and tracking}

We fit each point spread function with an ellipsoid to obtain the center of mass of the bulk and surface tracers (Fig.~\ref{fig_qd}C,D).
In this work, we measure local displacements by tracking the tracers between a set of initial and final configurations with the tracker described in \cite{Kim2021Measuring}.
We keep only the tracers found in all the confocal stacks.

In the bulk, we detect an average of 0.035 tracer per $\mu\mbox{m}^{-3}$, corresponding to a mean spacing of 3\,$\mu\mbox{m}$.
According to the supplier data sheet, the average mass fraction of quantum dots we dispersed in soft PDMS is $8\times 10^{-5}$.
With a density of about $6\,\mbox{g.cm}^{-3}$, this amounts for a volume fraction of $1.3\times 10^{-5}$, or 13 ppm.
From this we estimate that each bulk tracer is an aggregate of many quantum dots, with an average diameter of around 90 nm.
On the top surface, the 100\,nm polystyrene beads have an average number density of 0.1 tracer per $\mu\mbox{m}^{-2}$, also giving an inter-particle spacing of the order of 3\,$\mu\mbox{m}$. 
This corresponds to a surface occupation of roughly 800\,ppm.
Although a thorough study is left for future work, we here assume these fractions to be too small to influence mechanical properties. 

\subsection{A.4 Gradient calculation}
\label{sub:polyharmonic}

The fundamental quantity to measure material deformations is the gradient of displacements $\bm\nabla\mathbf u$, which we evaluate at each tracer location with a local polyharmonic spline interpolation scheme (see \cite{iske2004multiresolution}, Chapter 3.8).
It amounts to approximating at every position $\mathbf x = (x, y, z)$ the displacement field $u_\alpha(\mathbf x)$, which can be $u_x$, $u_y$, or $u_{\rm z}$, as the sum of $N$ radial base spline functions
\begin{equation}
	\tilde u_\alpha(\mathbf x) = \sum\limits_{i=1}^N w_i \varphi(\left\vert\mathbf x - \mathbf x_i\right\vert) + \mathbf v^{\text{\tiny T}}\cdot
	\left(
	\begin{array}{c}
	1\\
	x \\
	y \\
	z
	\end{array}
	\right),
	\label{eq:polyharmonic}
\end{equation}
where $\mathbf x_i$ are the positions of a subset of $N$ tracers, $w_i$ are weights to be determined, $\mathbf v = (v_0, v_x, v_y, v_z)$ a vector to be determined, and $\varphi(r) = r^3$ the radial base function.
We determine the unknowns $w_i$ and $\mathbf v$ by solving the linear system 
\begin{equation}
  \left(
  \begin{array}{cc}
  \mathbf A & \mathbf B\\
  \mathbf B^{\text{\tiny T}} & \mathbf 0
  \end{array}
  \right)
  \cdot
  \left(
  \begin{array}{c}
  \mathbf w \\
  \mathbf v
  \end{array}
  \right)
  =
  \left(
  \begin{array}{c}
  \mathbf u_{\alpha} \\
  \mathbf 0
  \end{array}
  \right),
  \label{eq:invert}
\end{equation}
where $A_{ij} = \varphi(\left\vert\mathbf x_i - \mathbf x_j\right\vert)$, $B_{ij} = \delta_{i1} + \delta_{i2} x_j + \delta_{i3} y_j + \delta_{i4} z_j$ with $\delta$ the is Kronecker symbol, and $u_{\alpha j}$ is the measured displacement of tracer $j$.

For the simple-shear experiments (\hyperref[sec:shear]{Appendix B}), we solve one linear system per tracer position $\mathbf x$.
For each tracer, we select a set of $N$ tracers such that $|x - x_i| < 10\,\mu$m, $|y - y_i| < 10\,\mu$m, and $|z - z_i| < 5\,\mu$m, and solve Eq.~\eqref{eq:invert} with the $N$ measured positions and displacements.
For the wetting ridge (\hyperref[sec:wet]{Appendix E}) and the coverslip edge measurements (\hyperref[sec:edge]{Appendix F}) we restrict the evaluation points to the surface tracers, and solve Eq.~\eqref{eq:invert} once using the $N$ tracers such that $z_i > -10$ microns.

As we chose the evaluation points to be tracers, we already know the displacements $u_\alpha(\mathbf x)$. We then estimate the associated gradients $\bm\nabla u_\alpha(\mathbf x)$ by deriving Eq.~\eqref{eq:polyharmonic}.

\begin{figure}[h!]
\includegraphics[scale=1]{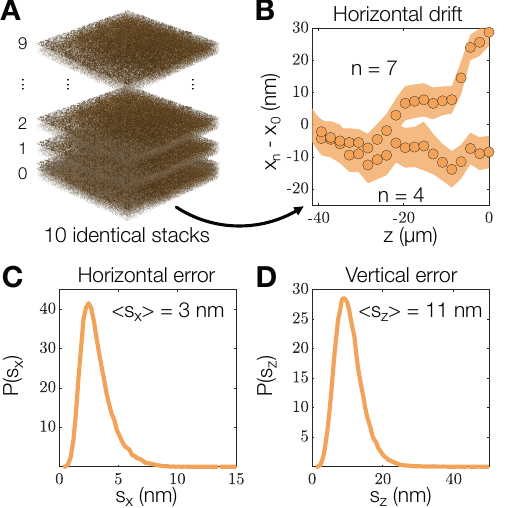}
\vspace{-10pt}
\caption{
(A) To determine the experimental noise, we image the same location ten consecutive times.
(B) Difference in the measured x positions between stacks 7 and 0, and between stacks 4 and 0. The circles and the size of the shaded areas respectively represent the average and twice the standard deviation within 2.2\,$\mu$m thick slices.
(C) Probability distribution function for the standard deviation of the horizontal detection.
(D) Probability distribution function for the standard deviation of the vertical detection.
Sketches not to scale.
}
\label{fig_meas}
\vspace{-15pt}
\end{figure}
\subsection{A.5 Measurement precision}

We take advantage of the static nature of our setup to quantify the precision of our measurements:
we image the same immobile location ten consecutive times, and compare the subsequently obtained positions (Fig.~\ref{fig_meas}A).
We slightly move the microscope stage relative to the objective to make sure that all measurements are uncorrelated.
We then detect the tracers in each stack, link them in all ten measurements with the tracker described in \cite{Kim2021Measuring}, and only keep the ones found in all ten stacks.
Instead of being identical, the obtained positions differ from one stack to the next.

With this procedure, we identify two sources of error.
A small z-dependent shift between each stack (Fig.~\ref{fig_meas}B), which we attribute to a slow drift of the microscope stage \cite{jonkman2020tutorial},
and an uncertainty in the evaluation of each point spread function's center of mass (Fig.~\ref{fig_qd}C).

\subsubsection{A.5.1 Stage drift}
\label{sec:drift}

\begin{figure}[h!]
\includegraphics[scale = 1]{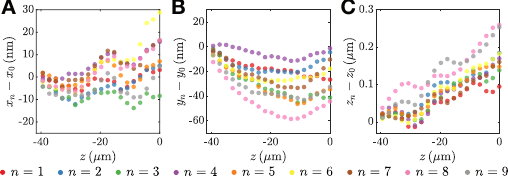}
\vspace{-10pt}
\caption{
Relative displacement between 10 consecutive stacks, in the (A) x direction, (B) the y direction, and (C) the z direction.
The color codes for the stacks numbering, and the circle for the mean value over a 2.2\,$\mu$m-thick slice.
}
\vspace{-15pt}
\label{fig_drift}
\end{figure}

In Fig.~\ref{fig_meas}B we show the difference in the measured x positions between stacks 7 and 0, and between stacks 4 and 0, and in Fig.~\ref{fig_drift} the relative displacement between the first stack and the nine other ones, in all $(x,y,z)$ spatial directions.
In all directions there is a net displacement which differs for each stack, without following any specific trend, of the order of tens of nanometers between the bottom and the top layers of the sample in the $(x, y)$ directions, and a few hundreds of nanometers in the z direction.

This behavior is consistent with a drift of the microscope stage.
If the microscope stage moves during the acquisition of the confocal stack, each slice will have an offset compared to its real position.
The difference in offset between two stacks then gives the relative displacement shown in Fig.~\ref{fig_drift}.
Here, we wait one second between each slice of the confocal stack, in order to let the immersive oil relax.
Imaging the whole stack then takes a few minutes, over which a $10$ to $100$ nm stage displacement is plausible \cite{jonkman2020tutorial}.

The consequence of this shift is a systematic error in the estimate of the displacement vertical gradients.
The largest horizontal shift being of $30$ nm over the $40\,\mu$m thick sample, it will result in a systematic error in the horizontal displacement vertical gradients $\partial u_x/\partial z$ and $\partial u_y/\partial z$ of the order of $0.1\%$.
Similarly, the largest vertical shift being of $250$ nm, it will result in a systematic error $\partial u_{\rm z}/\partial z \sim 0.5\%$ for vertical displacements.

\begin{figure*}
\includegraphics[width = \textwidth]{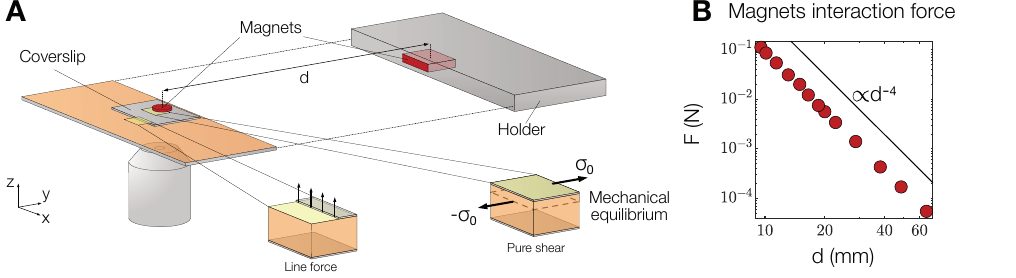}
\vspace{-15pt}
\caption{Simple shear experiment.
(A) Schematic of the bulk elasticity measurement setup.
At the edge of the coverslip, the interfacial tension acts as a line force (\hyperref[sec:edge]{Appendix F}).
Sketches not to scale.
(B) Intermagnet force as a function of intermagnet distance.
}
\vspace{-10pt}
\label{fig_mag}
\end{figure*}

\subsubsection{A.5.2 Location precision}

After correcting for stage drift, we obtain for each tracer a set of 10 estimated positions, for which we calculate the standard deviation in x, y, and z.
The probability distribution functions of these standard deviations resemble a Gamma distribution (Fig.~\ref{fig_meas}C,D), with an average standard error of $\langle s_x \rangle = 5$ nm in the $(x,y)$ directions and of $\langle s_z \rangle = 12$ nm in the $z$ direction, much smaller than the voxel dimensions $(108\times108\times270) \mbox{nm}^3$.
This leads to a random error in the displacement gradients of $\delta_{x,z} / L$, with $L$ the distance over which the gradient is calculated.
With a mean tracer-to-tracer distance of $3\,\mu\mbox{m}$, we estimate that in the absence of spatial averaging the error on the horizontal gradients are of $0.2\,\%$, and the error on the vertical gradients of $0.4\,\%$.

\section{Appendix B: Simple Shear}
\label{sec:shear}

In this section, we detail the simple-shear experimental setup, the magnet calibration, and we show displacement and tracer density profiles for all analysed samples not presented in the main text.

\subsection{B.1 Experimental protocol}

To conduct simple-shear measurements, we deposit a $1\times 1$ cm$^2$ coverslip on top of the soft PDMS, confining it between two parallel planes.
We then tape a small flat cylindrical magnet on the top coverslip, and mount a larger magnet on the sample holder, in a 3D-printed mobile part calibrated such that the two magnets are vertically aligned (Fig.~\ref{fig_mag}A).
By moving the mobile part, we can approach the two magnets to increase their attraction force, driving substrate deformation (Fig.~\ref{fig_mag}B).
While the large magnet is stuck in its holder, the small one is attached to the soft solid.
It then responds to the magnetic field by moving towards the larger magnet, shearing the material underneath.
In such geometry, mechanical equilibrium prescribes that the shear stress $\sigma_0$ is applied uniformly throughout the sample.\\

\noindent
In this geometry, we choose one location below the mobile coverslip and image it ten times.
This provides ten estimations of the reference configuration.
Then, each time we approach the magnets, we create a new deformed configuration.
We then image each deformed state twice and check there are no dynamic effects.

\subsection{B.2 Magnets calibration}

We estimate the magnets force-distance relationship with a precision scale.
We place the large magnet on the scale, standing on its edge, and gradually approach the vertically held smaller magnet.
We record the weight measured on the scale at various inter-magnet distances $d$.
The resulting force-displacement curve follows an expected dipole-dipole 4th order power law $F \propto d^{-4}$ (Fig.~\ref{fig_mag}B), which we fit to estimate the applied force at any separation distance $d$.

\subsection{B.3 z-dependent displacements}
\label{sec:z_dep_disp}

\begin{figure}[h!]
\includegraphics[scale = 1]{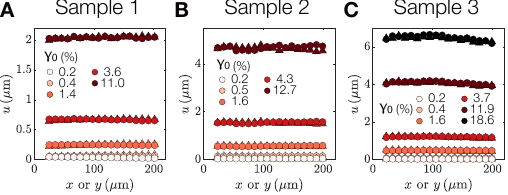}
\vspace{-10pt}
\caption{
Local horizontal displacements against $x$ (circles) and $y$ (triangles) for (A) sample 1, (B) sample 2 and (C) sample 3.
Symbols correspond to the displacement averaged over 10 $\mu$m wide slices.
}
\vspace{-15pt}
\label{fig_xy}
\end{figure}

We conducted shear measurements on all three samples we analysed.
In Fig.~\ref{fig_xy}, we show that the horizontal displacement $u$ is independent of $x$ and $y$ for all analysed samples, whatever the imposed strain.

\subsection{B.4 Non-linear displacement profiles}
\label{sec:shear_non_lin}

\begin{figure}[h!]
\includegraphics[scale = 1]{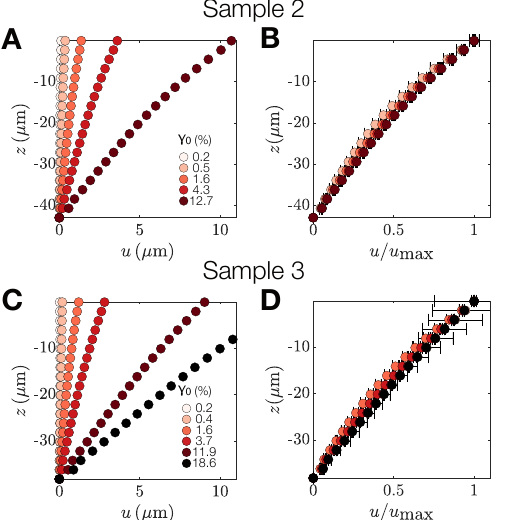}
\vspace{-10pt}
\caption{
Local horizontal displacements against vertical position $z$ for (A) sample 2 and (C) sample 3.
Normalized horizontal displacements against vertical position $z$ for (B) sample 2 and (D) sample 3.
Symbol positions and errorbars values computed as in Fig.~\ref{fig_bulk} (\hyperref[sec:err]{Appendix D}).
}
\vspace{-15pt}
\label{fig_shear}
\end{figure}

In Fig.~\ref{fig_bulk} we show that, in Sample 1, the horizontal displacement decays non-linearly with the distance from the surface.
The associated decaying profile is independent of the externally applied deformation, implying a heterogeneous shear modulus $\mu(z)$.
In Fig.~\ref{fig_shear}, we show that the two other analysed samples exhibit the same behavior.

\subsection{B.5 Origin of mechanical heterogeneity and tracer density profiles}
\label{sec:mech_hetero}

\begin{figure}
\includegraphics[scale = 1]{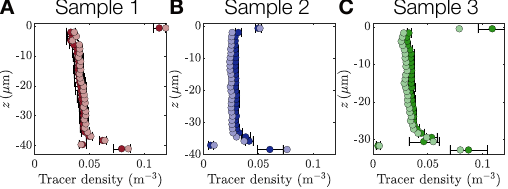}
\vspace{-10pt}
\caption{
Tracer density profiles for (A) sample 1, (B) sample 2, and (C) sample 3.
The light colors correspond to the density profile at the location on which the shear measurement is performed, and the dark colors represent the average density between all imaged locations of a single sample, and the errorbar twice the associated standard error of the mean.
}
\vspace{-15pt}
\label{fig_density}
\end{figure}

The physical origin of this heterogeneous response is unclear.
As a first test, we verify whether it is correlated to a heterogeneous tracer distribution within the material.
We estimate the tracer density as a function of vertical position by binning the z-axis in 20 slices, and dividing the number of tracers in each slice by the slice volume.
We show in Fig.~\ref{fig_density} the tracer density against the vertical position $z$ of each slice, for the three samples we analyzed.
For each sample, the density profile at the locations where the shear test was performed is similar to the density profile averaged over all imaged locations.
For all samples, there is an accumulation of tracers at the top and bottom surfaces, and a depletion zone close to the bottom surface.
For samples 1 and 3, there is a slight decrease in tracer density as approaching the surface, while the tracer density is homogeneous in the middle section for sample 2.

Variations in tracer density therefore do not explain the observed mechanical response.
Instead, we suspect it could arise either preparation details, such as toluene evaporation, or mixing heterogeneity during the spincoating process, or from crosslinking inhibituion by oxygen exposition of the top surface. Understanding these mechanisms is left for future work.

\section{Appendix C: Adhered coverslip}
\label{sec:coverslip}

As detailed in \hyperref[sec:shear]{Appendix B}, we add a small coverslip atop the thin PDMS film to confine it between two parallel plates and perform simple-shear tests (Fig.\ref{fig_mag}A).
We take advantage of this setup to complement the measurements at the three-phase contact line of sessile drops presented in the main manuscript.
At the edge of the coverslip, the soft PDMS is deformed by capillary effects.
It thus provides a deformed air-PDMS interface, that has not been in contact with a liquid drop, and which we can additionally deform by exerting a force on the magnet attached to the top coverslip.
Results of these measurements are presented in \hyperref[sec:edge]{Appendix F}.
\begin{figure*}
\includegraphics[width = \textwidth]{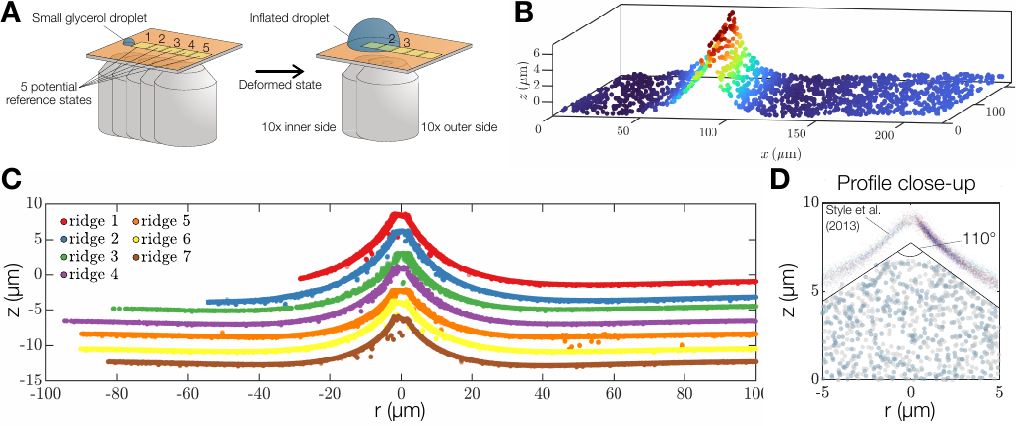}
\vspace{-10pt}
\caption{
(A) Sketch of the acquisition of reference and deformed configurations for the measurements close to the three-phase contact line.
(B) 3D  profile for the wetting ridge induced by a glycerol sessile drop. Each point correspond to a detected surface tracer, and the color codes for the z position.
(C) Azimuthally collapsed surface tracer positions for all the imaged wetting ridges. Each color gathers ten distinct  measurements of the tracer positions for a single wetting ridge (see Table \ref{table_ridge} for a list of all the measured ridges). The ridges are vertically shifted with respect to each other for visualisation purposes.
(D) Azimuthally-collapsed tracer positions on both sides of the wetting ridge (same as Fig.~\ref{fig_droplet}B, different scale). The cloud of points above the solid lines corresponds to the wetting ridge data presented in \cite{style2013universal}, to scale.
}
\vspace{-15pt}
\label{fig_ref_state}
\end{figure*}

\section{Appendix D: Error estimation}
\label{sec:err}
We systematically estimate the standard errors from sets of repeated measurements.

For the simple-shear tests, we image ten reference states, and two deformed states for each magnet position (Fig.~\ref{fig_mag}A).
For each reference state, for every tracer, we calculate the local horizontal displacement $u$ and strain $\partial u / \partial z$ with respect to the two deformed states, giving 20 measurements for each deformation state.
We then bin the z-axis in 2.2\,$\mu$m thick slices, and compute the mean displacement and strain within each bin.
The plotted values and errorbars in Fig.~\ref{fig_bulk} respectively correspond to the average and twice the standard deviation of the 20 mean values for each bin.

For the measurements at the wetting widge of a sessile drop or at the edge of a coverslip, we image one reference state, and ten deformed states (Fig.~\ref{fig_ref_state}A).
For every tracer, we calculate the local displacements $\mathbf u$ and strains $\mathbf E$ between the initial state and each deformed states, giving 10 measurements.
We then bin the r-axis in 5\,$\mu$m wide segments, and compute the mean displacement and strain within each bin.
The plotted values and errorbars in Fig.~\ref{fig_droplet} and Fig.~\ref{fig_jump} respectively correspond to the average and twice the standard deviation of the 10 mean values in each bin.

\section{Appendix E: Wetting ridge-induced deformations}
\label{sec:wet}
In this section, we provide more details about the measurements close to the wetting ridge of a sessile drop.
We explain how we measure the undeformed state, we show a typical 3D profile, we detail the calculation of surface unit vectors, and we provide surface strain and surface tension measurements for all analysed samples.

\subsection{E.1 Wetting ridge reference state}
\label{sec:ridge_ref_state}

To obtain undeformed, reference states, we proceed as depicted in Fig.~\ref{fig_ref_state}A.
First, we deposit a small droplet of glycerol on the soft silicone gel.
Second, we acquire 5 overlapping confocal stacks close to the droplet, where the material is undeformed, covering a distance of one millimeter.
These constitute a pool of potential reference states.
Next, we expand the small droplet with extra glycerol, until its contact line reaches the first imaged undeformed state.
We then let the droplet equilibrate overnight, and the location where the contact line stopped indicates which reference states to use.
At last, we image 10 times the deformed states, one on each side of the contact line, each corresponding to one of the acquired reference states.

\subsection{E.2 Wetting ridge profile}
\label{sec:ridge_profile}

In Fig.~\ref{fig_ref_state}B, we show a typical 3D profile for the wetting ridge created by a sessile drop on a soft silicone solid.
In the main manuscript, we exploit the azimuthal symmetry of the profile.
Importantly, we note that for any given wetting ridge, the surface tracer azimuthally-collapsed positions of the ten consecutive confocal stacks are identical (Fig.~\ref{fig_ref_state}C).
This ensures that the system is at mechanical equilibrium, and that no motion occurs at the time scale of the experiment.

Although previous work found reported a sharp wetting ridge profile \cite{style2013universal},  the wetting ridges here are truncated
as in \cite{cai2021fluid} (Fig.~\ref{fig_ref_state}C,D).
This discrepancy is likely due to the differences in waiting time between droplet deposition and imaging.  
In \cite{style2013universal}, the authors image the wetting ridge immediately after depositing a droplet.
Here, we do so after several hours.
During that time, the high stresses at the tip of the ridge expel some of the free chains, and the geometry relaxes poroelastically to its equilibrium shape \cite{flapper2023reversal, park2014visualization}.
Nevertheless, the measured opening angle of $110^\circ$ is consistent with the one observed at that scale in \cite{style2013universal} (Fig. 4C).

\subsection{E.3 Tangential and normal vectors}
\label{sec:tang_norm}

We estimate the surface tangential and normal unit vectors from the radially collapsed tracer positions in the deformed state $(r, z)$, which correspond to the surface points of Fig.~\ref{fig_droplet}B.

We smooth the vertical positions with the Matlab \textit{csaps} function to $z_{\rm sm}$, we compute the radial and vertical gradients $(\mbox{d}r, \mbox{d}z_{\rm sm})$ with the Matlab \textit{gradient} function, and the radial tangent and normal vectors $\mathbf t_{\rm r} = (\mbox{d}r, \mbox{d}z_{\rm sm}) / \sqrt{\mbox{d}r^2 + \mbox{d}z_{\rm sm}^2}$, $\mathbf  n_{\rm r} = (-t_{\rm r}(2) , t_{\rm r}(1))$.
We then project them on the orthoradial line associated to the angular position $\theta$ of each tracer to obtain the 3D tangent and normal vectors in the cartesian coordinate system $\mathbf t = (t_{\rm r}(1) \cos{\theta}, t_{\rm r}(1)\sin{\theta}, t_{\rm r}(2))$, $\mathbf n = (n_{\rm r}(1) \cos{\theta}, n_{\rm r}(1)\sin{\theta}, n_{\rm r}(2))$.
\begin{figure}[h!]
\includegraphics[scale = 1]{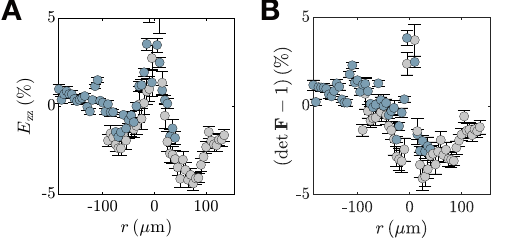}
\vspace{-10pt}
\caption{
(A) Surface vertical strain $E_{\rm zz}$ and (B) surface volumetric strain $(\det\mathbf F - 1)$.
Same dataset as in the main text. The symbols correspond to values averaged over 5 $\mu$m wide radial bins, the size of the errorbars equate twice the standard error of the mean (\hyperref[sec:err]{Appendix D}). The blue and gray colors correspond to confocal datasets with most of the field of view on the glycerol and air sides, respectively.
}
\vspace{-15pt}
\label{fig_norm_vol_surf}
\end{figure}

\subsection{E.4 Surface normal and volumetric strains}
\label{sec:surf_nom_vol}

For sake of completeness, we here report the surface vertical strain $E_{\rm zz}$ (Fig.\ref{fig_norm_vol_surf}A).
In contrast with the surface strains presented in the main text, we note a small vertical shift between the normal strains acquired with the field of views mostly on the air side (gray) and the ones mostly on the glycerol side (blue).
This shift is consistent with the stage drift reported in \hyperref[sec:drift]{Appendix A.5.1}: we estimate the displacement vertical derivatives up to an unknown offset, corresponding to the difference in stage drift between the reference and deformed states, about five times larger for $\partial u_{\rm z} / \partial z$ ($\sim0.5\%$) than for $\partial u_x / \partial z$ and $\partial u_y / \partial z$ ($\sim0.1\%$).
We therefore interpret the vertical shift in normal strains (Fig.\ref{fig_norm_vol_surf}A and B) as a result of their first-order dependence on $\partial u_{\rm z} / \partial z$, which the strains presented in the main text do not have.

Similarly, we report the surface volumetric strains $\det\mathbf F$(Fig.\ref{fig_norm_vol_surf}B).
They also depend to first order on $\partial u_{\rm z} / \partial z$, and therefore also present a vertical shift between the strains acquired with the field of views mostly on the air side (gray) and the ones mostly on the glycerol side (blue).
In addition, we note that the surface volumetric strains are non-zero, a signature of poroelastic deformations.
Such signature is not suprising for silicone gels, which contain a large volume fraction of free polymeric chains \cite{flapper2023reversal,xu2020viscoelastic,park2014visualization}.

\subsection{E.5 Bulk volumetric strains}
\label{sec:vol_bulk}

\begin{figure}[h!]
\includegraphics[scale = 1]{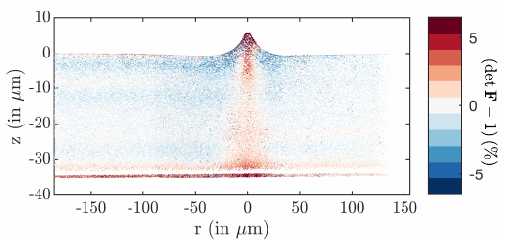}
\vspace{-10pt}
\caption{
Azimuthally-collapsed tracer positions, colored following the estimated volumetric strain $\det\mathbf F - 1$ values, for the dataset presented in the main text. The confocal datasets corresponding to the field of views on the glycerol and air sides are superimposed. For visualisation purposes, a vertical rescale of about $102\%$ is applied to the confocal stack on the air side to compensate for the stage drift and ensure a good collapse of the tracer positions.
}
\label{fig_vol_bulk}
\end{figure}

We provide an overview of the poroelastic deformations under a sessile drop by estimating the volumetric strains at all the tracer positions (Fig.~\ref{fig_vol_bulk}).
In agreement with recent work \cite{flapper2023reversal}, the region below the acute wetting ridge swells, with a degree of swelling that is maximal at the ridge tip. 
In contrast, however, this swelling extends way beyond the vicinity of the wetting ridge. 
Although the absolute swelling values have to be interpreted with care, because of the uncertainty resulting from the stage drift, the long-range swelling below the wetting ridge is significant and its detailed investigation is left for future work.

\newpage
\subsection{E.6 Surface elasticity measurements in all samples}
\label{sec:surf_el_all_samples}

We performed the surface elasticity measurements described in the main text on multiple locations of the wetting ridge, on three different samples.
On Sample 1, we imaged the PDMS-air interface at two locations, on Sample 2 the PDMS-air and the PDMS-glycerol interfaces at two locations, and on Sample 3 the PDMS-air and the PDMS-glycerol interfaces at three locations.
The results presented in the main manuscript were performed on one of the ridges of Sample 3.
All measurements and their specifics are presented in Table~\ref{table_ridge}.

\begin{table}[h]
\centering
\begin{tabularx}{\columnwidth}{c|c|c}
Name&Sample \#&Specifics\\
 \hline\hline
 Ridge 1 & 1 & PDMS-air interface\\
 Ridge 2 & 1 & PDMS-air interface \\
 Ridge 3 & 2 & PDMS-air and glycerol interfaces\\
 Ridge 4 & 2 & PDMS-air and glycerol interfaces\\
 Ridge 5 & 3 & PDMS-air and glycerol interfaces (main)\\
 Ridge 6 & 3 & PDMS-air and glycerol interfaces\\
 Ridge 7 & 3 & PDMS-air and glycerol interfaces\\
 \hline\hline
\end{tabularx}
\caption{Summary table for the wetting ridge measurements.
Ridge 5 corresponds to the measurement presented in the main manuscript.}
\label{table_ridge}
\end{table}

\begin{figure}[h!]
  \includegraphics[scale = 1]{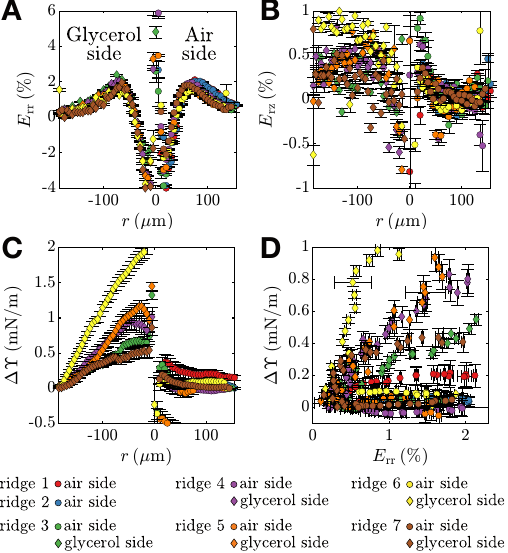}
\vspace{-10pt}
\caption{
  (A) and (B) Surface radial stretch $E_{\rm rr}$ and vertical shear stretch $E_{\rm rz}$.
  (C) and (D) Interfacial tension increment $\Delta \Upsilon$ against distance from the contact line $r$, and against radial stretch $E_{\rm rr}$.
  Each color corresponds to a different measurement, the circles (resp. diamonds) correspond to values on the air side (resp. on the glycerol side) averaged over 5 $\mu$m wide radial bins, and the size of the errorbars equates twice the standard error of the mean.
  }
\vspace{-15pt}
\label{fig_ridges}
  \end{figure}

For each measurement, we computed the radial stretch $E_{\rm rr}$, the vertical shear stretch $E_{\rm rz}$, and the associated surface tension increment $\Delta\Upsilon$ (Fig.~\ref{fig_ridges}) on the sample surface.
The radial stretches $E_{\rm rr}$ induced by the presence of the sessile drop all collapse on the same curve (Fig.~\ref{fig_ridges}A).
As described in the main text, the radial stretch increases from 0 in the far-field to 2\% in the vicinity of the contact line, and is symmetric between the side that is beneath the glycerol droplet, and the side that is not.

Similarly, all measures of the vertical shear stretch $E_{\rm rz}$ fall onto each other.
They are, however, no longer symmetric with respect to the contact line.
The vertical shear stretch averages to zero in the air side of the interface, and  is non-zero on the glycerol side (Fig.~\ref{fig_ridges}B).
As a consequence, the surface tension increment $\Delta\Upsilon$ remains null on the air side, while it increases as it gets closer to the contact line on the glycerol side (Fig.~\ref{fig_ridges}C).
When compared to the radial stretch, this difference systematically amounts to a stretch-independent surface tension at the air-PDMS interface, and a stretch-dependent surface tension at the glycerol-PDMS interface (Fig.~\ref{fig_ridges}D).


We performed a linear regression for the surface tension increment evolution with stretch (Fig.~\ref{fig_ridges}D), restricted to the surface points resting further than 65\,$\mu$m from the wetting ridge to stay in the small deformation regime, and gathered the measured elastic moduli in Table~\ref{table_ridge_Lambda}.
Overall, we obtain an average surface elastic modulus of $\Lambda_{\rm air} = 1.6 \pm 1.7$~mN/m for the air-PDMS interface, and of $\Lambda_{\rm gly} = 45.2 \pm 21.6$~mN/m for the glycerol-PDMS interface.

  \begin{table}[h]
    \centering
    \begin{tabularx}{\columnwidth}{c|l|l}
    \hline\hline
    Name&\multicolumn{2}{c}{Surface elastic modulus (in mN/m)}\\
     \hline
     Ridge 1 &$\Lambda_{\rm air} = 3.0\pm 2.2$&\\
     Ridge 2 &$\Lambda_{\rm air} = 3.0\pm 0.6$&\\
     Ridge 3 &$\Lambda_{\rm air} = 2.9 \pm 0.6$&$\Lambda_{\rm gly} = 24.3 \pm 1.9$\\
     Ridge 4 &$\Lambda_{\rm air} = -1.4 \pm 0.6$&$\Lambda_{\rm gly} = 43.2 \pm 2.5$\\
     Ridge 5 &$\Lambda_{\rm air} = 1.6 \pm 2.2$&$\Lambda_{\rm gly} = 48.1 \pm 4.4$ (main)\\
     Ridge 6 &$\Lambda_{\rm air} = 0.2 \pm 0.6$&$\Lambda_{\rm gly} = 80.0 \pm 18.2$\\
     Ridge 7 &$\Lambda_{\rm air} = 2.1 \pm 0.3$&$\Lambda_{\rm gly} = 30.7 \pm 5.0$\\
    \hline\hline
    \end{tabularx}
    \caption{Summary table of the estimated surface elastic moduli for the wetting ridge measurements.
    Ridge 5 corresponds to the measurement presented in the main text.
    Error bars correspond to 95\% confidence intervals, obtained from a linear regression.}
    \label{table_ridge_Lambda}
    \end{table}
    
    \section{Appendix F: Edge-induced deformations}
    \label{sec:edge}
    
\begin{figure}
      \includegraphics[scale = 1]{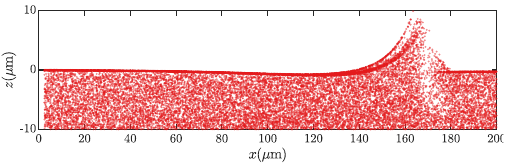}
      \vspace{-10pt}
      \caption{
      Tracer $(x,z)$ positions at the edge of the coverslip for Sample 1 after coverslip deposition.
      }
      \vspace{-15pt}
      \label{fig_edge}
      \end{figure}

    To complement the measurements presented in the main text, we imaged the displacements at the PDMS-air interface on the edge of the deposited coverslip (Fig~\ref{fig_mag}A).
    There, the interfacial tension between the coverslip and the silicone gel creates a wetting ridge, where the substrate is deformed (Fig.~\ref{fig_edge}).
    In addition, displacing the coverslip further deforms the silicone gel at its edges, allowing for more control on the imposed deformations.

To estimate the resulting strains and stresses, we follow a procedure similar to the one sketched in Fig.~\ref{fig_ref_state}A.
We imaged a series of potential reference states, deposited the coverslip, let the system relax overnight, and imaged the deformed state 10 times.
We then displaced the top coverslip by changing the inter-magnet distance, kept it still overnight, and imaged the new deformed state 10 times.
In total, we obtained eight different configurations, belonging to Sample 1 and Sample 3 (a piece of dust unfortunately landed on the field of view of Sample 2), with specifics detailed in Table~\ref{table_edges}.

    \begin{table}[h]
    \centering
    \begin{tabularx}{\columnwidth}{c|c|c}
    \hline\hline
    Name&Sample \#&Specifics\\
     \hline
     Edge 1 & 1 & after coverslip deposition\\
     Edge 2 & 1 & after coverslip displacement\\
     Edge 3 & 3 & location 1, after coverslip deposition\\
     Edge 4 & 3 & location 1, after coverslip displacement 1\\
     Edge 5 & 3 & location 1, after coverslip displacement 2\\
     Edge 6 & 3 & location 2, after coverslip displacement 1\\
     edge 7 & 3 & location 2, after coverslip displacement 2\\
     \hline\hline
    \end{tabularx}
    \caption{Summary table for the edge measurements.}
    \label{table_edges}
    \end{table}
      
\begin{figure}[h!]
      \includegraphics[scale = 1]{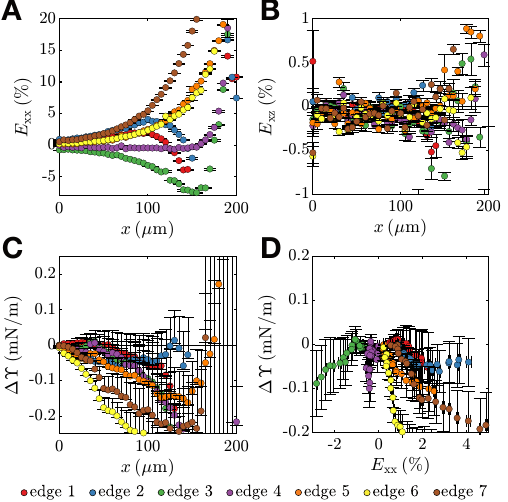}
      \vspace{-10pt}
      \caption{
      (A) and (B) Surface stretch $E_{\rm xx}$ and vertical shear stretch $E_{\rm xz}$.
      (C) and (D) Interfacial tension increment $\Delta \Upsilon$ against the longitudinal position $x$, and against longitudinal stretch $E_{\rm xx}$.
      Each color corresponds to a different measurement, the circles correspond to values averaged over 5 $\mu$m wide longitudinal bins, and the size of the errorbars equate twice the standard error of the mean.
      }
      \vspace{-15pt}
      \label{fig_edges}
      \end{figure}

      For every sample, we placed the coverslip such that its edges were aligned with the $(x, y)$ axis of the microscope stage.
      This way, the deformation is nearly symmetric along the $y$ direction, and we only consider variations along the $x$ direction.
      Similarly to the measurements below the sessile drops presented in the main text, we compute the stretch of the surface points $\mathbf E = (\mathbf F^{\text{\tiny T}}\mathbf F - \mathbf I) /2$.
      The surface strain in the longitudinal direction is equal to $E_{\rm xx} = \hat{\mathbf x}\cdot\mathbf E\cdot\hat{\mathbf x}$ (Fig.~\ref{fig_edges}A), with $\hat{\mathbf x}$ the unit vector in the $x$ direction, and the vertical shear stretch to $E_{\rm xz} = \hat{\mathbf x}\cdot\mathbf E\cdot\hat{\mathbf z}$ (Fig.~\ref{fig_edges}B).
    
      When the coverslip is displaced, from edges 1 to 2, edges 3 to 4 to 5, and edges 6 to 7, the longitudinal stretch systematically increases (Fig.~\ref{fig_edges}A).
      The longitudinal stretch just after deposition, however, varies greatly between the different measures.
      The surface of edge 1 is in tension, the one of edge 3 in compression, and the one of edge 4 nearly undeformed (Fig.~\ref{fig_edges}A).
      Probing the deformation near the edge of the coverslip therefore allows to test for different stretch conditions.
      In contrast to the longitudinal stretch $E_{\rm xx}$, the vertical shear stretch $E_{\rm xz}$ remains systematically close to zero, with no net average (Fig.~\ref{fig_edges}B).
      Consequently, the surface tension increment also remains small (Fig.~\ref{fig_edges}C), and no surface elasticity is visible in any of the measurements (Fig.~\ref{fig_edges}D).
      This result suggests that, independently of the deformation state, the surface tension at the PDMS-air interface remains constant.
      
      To estimate the elastic moduli of the air-PDMS interfaces at the vincinity of the edge of the coverslip, we performed a linear regression for the surface tension increment evolution with stretch (Fig.~\ref{fig_edges}E), restricted to the surface points resting further than 100\,$\mu$m from the edge to stay in the small deformation regime, and gathered the measured elastic moduli in Table~\ref{table_edges_Lambda}.
      Overall, we obtain a surface elastic modulus of $\Lambda_{\rm air} = -2.2 \pm 7.1$~mN/m.
      
      \begin{table}[h]
      \centering
      \begin{tabularx}{\columnwidth}{c|c}
       \hline\hline
      Name&Surface elastic modulus (in mN/m)\\
       \hline
       Edge 1 & $\Lambda_{\rm air} = -1.2\pm 0.5$\\
       Edge 2 & $\Lambda_{\rm air} = -1.5\pm 0.3$\\
       Edge 3 & $\Lambda_{\rm air} = 2.2\pm 0.5$\\
       Edge 4 & $\Lambda_{\rm air} = 8.9\pm 3.4$\\
       Edge 5 & $\Lambda_{\rm air} = -5.1\pm 0.6$\\
       Edge 6 & $\Lambda_{\rm air} = -14.3\pm 1.6$\\
       Edge 7 & $\Lambda_{\rm air} = -4.2\pm 0.4$\\
       \hline\hline
      \end{tabularx}
      \caption{Summary table of the estimated surface elastic moduli for the edge measurements.
      Error bars correspond to 95\% confidence intervals, obtained from a linear regression.
      }
      \label{table_edges_Lambda}
      \end{table}
      
Overall, we find a net solid-like response of the glycerol-PDMS interface.
The five independently measured moduli are all significantly positive, and range from $\Lambda = 24.3~\mathrm{mN/m}$ to $\Lambda = 80.0~\mathrm{mN/m}$, with a statistical error lower than 20\%.
By contrast, over fourteen independent measurements, we obtain a virtually liquid-like response of the air-PDMS interface, with moduli ranging from $\Lambda = -14.3~\mathrm{mN/m}$ to $\Lambda = 8.9~\mathrm{mN/m}$.

\section{Appendix G: Bulk-extrapolated displacements and strains}
\label{sec:bulk_extr}

To obtain the bulk-extrapolated surface displacement and strains, we proceed as follows.
From the set of measured displacements $\mathbf u_i = \mathbf u(\mathbf x_i)$, where $\mathbf x_i$ corresponds to all tracer positions, we remove the surface tracers $\mathbf x_{i_s}$.
We thus obtain another pool of measured displacements  $\mathbf u_{i_b} = \mathbf u(\mathbf x_{i_b})$, where $\mathbf x_{i_b}$ corresponds to all tracer positions located in the bulk of the material.
We then solve the linear system Eq.~\eqref{eq:invert} using the $N$ bulk tracers such that $z_{i_b} > -10\,\mu$m.
At last, we plug the obtained $\mathbf v$ and $\mathbf w$ vectors into Eq.~\eqref{eq:polyharmonic} and its spatial derivatives, and use them to calculate bulk-extrapolated surface displacements and strains at all the surface point locations $\mathbf x = \mathbf x_{i_s}$.

\section{Appendix H: Displacement jump and surface elasticity}
\label{sec:disp_jump_app}

\begin{figure}[h!]
  \includegraphics[scale = 1]{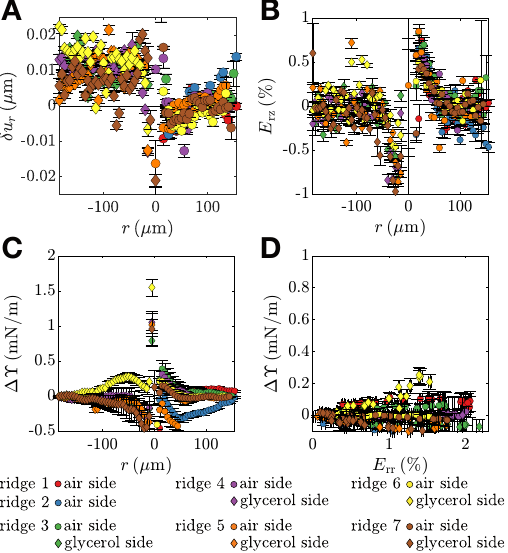}
  \caption{
  (A) and (B) Surface radial displacement jump $\delta u_{\rm r}$ and bulk-extrapolated vertical shear stretch $E_{\rm rz}$.
  (C) and (D) Bulk-extrapolated interfacial tension increment $\Delta \Upsilon$ against distance from the contact line $r$, and against radial stretch $E_{\rm rr}$.
  Each color corresponds to a different measurement, the circles (resp. diamonds) correspond to values on the air side (resp. on the glycerol side) averaged over 5 $\mu$m wide radial bins, and the size of the errorbars equates twice the standard error of the mean.
  }
  \label{fig_ridges_no_surf_pts}
\end{figure}

In Fig.~\ref{fig_ridges_no_surf_pts}, we show the displacement jump and the bulk-extrapolated shear strain and resulting interfacial tension for all the analysed samples.
The displacement jump is systematically positive on the glycerol side, and close to zero on the air side (Fig.~\ref{fig_ridges_no_surf_pts}A).
When we extrapolate the bulk information to the surface, the shear strain at the surface becomes symmetric between both sides of the wetting ridge (Fig.~\ref{fig_ridges_no_surf_pts}B), and the interfacial tension looses its strain dependence on the PDMS-glycerol interface (Fig.~\ref{fig_ridges_no_surf_pts}C,D).
These results convincingly show that the presence of a displacement jump at the interface is responsible for the measured surface elasticity.

\begin{figure*}%
\includegraphics[scale = 1]{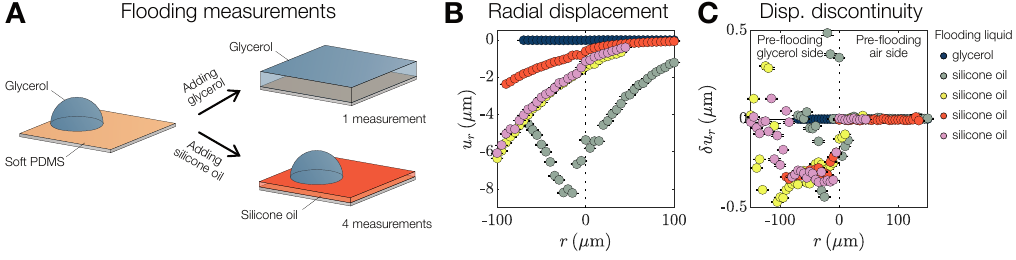}
\caption{
(A) Sketch of the flooding tests. After the measurements on the wetting ridge, in one instance we flooded the system with glycerol, and in four instances with silicone oil. (B) Surface radial displacements and (C) surface radial displacement discontinuity, against radial position $r$.
Dark blue circles correspond to a glycerol-flooded sample, and other colors to different silicone-flooded samples.
The $r = 0\,\mu\mbox{m}$ position corresponds to the pre-flooding wetting ridge location, positive $r$ to the pre-flooding PDMS-air interface, and negative $r$ to the pre-flooding PDMS-glycerol interface.
}
\label{fig_gly_sil}
\end{figure*}

\section{Appendix I: Displacement jump is not an optical artifact}
\label{sec:dis_jump_not_artifact}

A priori, we can rule out the possibility of an optical artifact for two reasons.
First, imaging was performed with an inverted microscope.
The collected light only goes through the immersion fluid and the PDMS film.
Its travel path is therefore identical for the surface tracers at the PDMS-air and PDMS-glycerol interfaces.
Second, if there was a refractive index change over a small thickness below the PDMS-glycerol interface, the vertical displacement $u_{\rm z}$ would be affected but not the radial displacement $u_{\rm r}$.

For sake of completeness, we proceeded to two additional tests.
In one instance we flooded the PDMS surface with glycerol, and in four instances we flooded it with silicone oil (Gelest DMS-V35) (Fig.~\ref{fig_gly_sil}A).
We then compute the tracer displacement as the difference between their position in the undeformed configuration (Fig.~\ref{fig_ref_state}A) and the flooded configuration.
For clarity, we azimuthally collapse the data according to the pre-flooding configuration.
This way, the $r=0\,\mu\mbox{m}$ position corresponds to the location of the wetting ridge prior to flooding.
In Fig.~\ref{fig_gly_sil}B we show the surface radial displacement for surfaces flooded with glycerol and silicone oil, and in Fig.~\ref{fig_gly_sil}B the difference with their bulk-extrapolated equivalent.

For the glycerol-flooded surface, the surface is nearly unstretched (Fig.~\ref{fig_gly_sil}B), meaning that the surface tracers went back to their initial position.
This observation shows that the surface did not undergo irreversible deformation, and that the surface tracers accurately follow the surface motion.
Furthermore, the radial displacements present no measurable discontinuity (Fig.~\ref{fig_gly_sil}C).

For the silicone-immersed surfaces, the wetting ridge moves with the addition of silicone oil and the position of the new wetting ridge is located in negative $r$ values.
Although the amplitude of this motion varies between the different tests, it remains close enough to its previous location to induce radial displacements within the pre-flooding field of view (Fig.~\ref{fig_gly_sil}B).
For all four silicone-immersed surfaces, there are no measurable jump in the radial displacement at positive $r$ values (Fig.~\ref{fig_gly_sil}C).
There, the PDMS surface is in contact with silicone oil, and was in contact with air prior to flooding.
At negative $r$ values, we however systematically observe a non-zero jump in the radial displacement (Fig.~\ref{fig_gly_sil}C).
There, the PDMS surface is in contact with silicone oil, but was in contact with glycerol prior to flooding.

Together, these results demonstrate that the presence of a jump in radial displacement at the PDMS surface cannot be uniquely attributed to the nature of the fluid in contact with it.
Instead, two conditions are required for this displacement jump in our measurements: the surface must be stretched, and it must have experienced the passage of a glycerol wetting ridge.
This decisively rules out the hypothesis of an optical artifact and suggests a permanent alteration of the near-surface mechanical properties.

\section{Appendix J: Swelling of PDMS by glycerol}
\label{sec:glycerol}

\begin{figure}[h!]
\includegraphics[scale = 1]{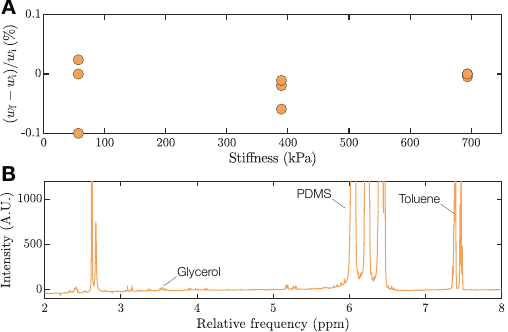}
\caption{
(A) Glycerol relative mass intake for samples of three different stiffness.
(B) NMR spectrum for liquid PDMS in which glycerol diffused for two weeks.
}
\label{fig_glycerol}
\end{figure}

To check whether glycerol swells PDMS gels, we conducted two separate measurements.
The first consisted in immersing a piece of PDM gel of diameter 10 mm, thickness 2.8 mm, and initial weight $w_{\rm i}\sim200$ mg, into a bath of glycerol, let it soak for 48 hours at room temperature, and measure the final weight $w_{\rm f}$ after wiping the sample with a tissue to remove the excess glycerol on the surface.
The silicone gels were made by mixing silicone chains (Gelest DMS-V31, 28 kDa), a crosslinker (Gelest HMS-301), and a catalyst (Gelest SIP6831.2), with elastic modulus varying with the ingredients ratio. 
We tested three stiffness values, each three times.
For all measured sample, the gylcerol relative mass intake $(w_{\rm f} - w_{\rm i}) / w_{\rm i}$ averages around $0\%$, with a dispersion smaller than $0.1\%$ (Fig.~\ref{fig_glycerol}A).
showing a negligible swelling of the silicone gel by glycerol.

The second is a NMR measurement. 
We place glycerol in a funnel that is plugged at the bottom with a stopper, and add liquid PDMS above. 
We carefully handle this operation, such that no mechanical mixing occurs and the only form of mixing is through diffusion.
We let this setup stand still for 2 weeks, before we remove most of the glycerol phase from the funnel by taking the stopper off. 
We then carefully pipette out 1 mL of the liquid PDMS phase, as close as possible ($\sim$ 1mm) to the glycerol boundary, while making sure not to aspirate the glycerol phase.
We mix this sample with toluene to reach a concentration of 1mM, and transfer it to a Novell S500 NMR tube.

We determined the glycerol concentration using proton nuclear magnetic resonance
($^1$H NMR, Bruker AV-500) spectroscopy \cite{testa2021sustained}, with toluene as a solvent (Fig.~\ref{fig_glycerol}B).
We perform 512 scans, with a relaxation delay of 3 s.
As we know the concentration of toluene, we deduce the concentration of glycerol by comparing the area under the glycerol peak ($[3.46, \,3.64]$ ppm) to the one under the toluene peak ($[7.31, \,7.52]$ ppm), while accounting for the number of hydrogen atoms in both toluene and glycerol molecules.
This gives a glycerol concentration of $0.03$ mM.
With a molar concentration of $92$ g/mol, it leads to a 3 ppm concentration of glycerol into PDMS.

\section{Appendix K: Supplementary Videos}
\label{sec:supp_videos}

\textbf{Supplementary Video 1:} Confocal stack for an undeformed PDMS sample seeded with quantum dots.
Each frame corresponds to a horizontal slice, each separated from the next by a vertical shift of 250 nm.
\\

\textbf{Supplementary Video 2:} Diffusion of 200 nm fluorescent beads dispersed into a glycerol droplet sitting on a soft silicone.
The films are taken at the contact line.
The left-hand-side movie corresponds to a focal plane close to the bottom surface, and the right-hand-side to a focal plane 10 $\mu$m above the bottom surface.
Two successive frames are separated by a time step of 500 ms.

To create a shear stress $\sigma_{\rm tn} \sim 3\,\mbox{Pa}$, comparable to the one resulting to the stretch measured under the glycerol (Fig.~\ref{fig_ridges}B), the glycerol would need to exhibit a shear flow of strain rate $\partial v / \partial z = \sigma_{\rm tn} / \eta \sim 2\,\mbox{s}^{-1}$, where $v$ is the glycerol velocity parallel to the bottom surface and $\eta$ the glycerol viscosity.
At 10 $\mu$m above the surface, the glycerol velocity should therefore be of the order of 20 $\mu$m/s.
In Supplementary Video 2, we do not see any sign of such a drift.
Instead, the nanotracers dispersed inside the glycerol diffuse without any visible mean flow, showing that the observed shear strain at the glycerol-PMDS interface cannot be explained by fluid flows inside the glycerol.

\renewcommand\refname{Bibliography}
\bibliographystyle{apsrev4-1} 
\bibliography{paper_1_biblio} 

\end{document}